\documentclass[12pt,epsfig]{article}

\newcommand{\beq}{\begin{equation}}
\newcommand{\eeq}{\end{equation}}
\newcommand{\beqa}{\begin{eqnarray}}
\newcommand{\eeqa}{\end{eqnarray}}

\newtheorem{theorem}{Theorem}
\newtheorem{lemma}{Lemma}

\begin{document}

\def\rl{\rangle \langle}
\def\openone{\leavevmode\hbox{\small1\kern-3.8pt\normalsize1}}
\def\RR{{\rm I\kern-.2emR}}
\def\tr{{\rm tr}}
\def\ce{{\cal E}}
\def\cc{{\cal C}}
\def\ci{{\cal I}}
\def\cd{{\cal D}}
\def\cb{{\cal B}}
\def\cn{{\cal N}}
\def\ct{{\cal T}}
\def\cf{{\cal F}}
\def\ca{{\cal A}}
\def\cg{{\cal G}}
\def\cv{{\cal V}}
\def\cc{{\cal C}}
\def\rhon{\rho^{\otimes n}}
\def\on{^{\otimes n}}
\def\pn{^{(n)}}
\def\pnp{^{(n)'}}
\def\tcd{\tilde{\cal D}}
\def\tcn{\tilde{\cal N}}
\def\tct{\tilde{\cal T}}
\def\id{\frac{I}{d}}
\def\pthang{\frac{P}{\sqrt{d \tr P^2}}}
\def\psirq{\psi^{RQ}}
\def\rhorq{\rho^{RQ}}
\def\rhorqp{\rho^{RQ'}}
\def\ra{\rangle}
\def\la{\langle}
\def\QED{\mbox{\rule[0pt]{1.5ex}{1.5ex}}}
\def\proof{\noindent\hspace{2em}{\it Proof: }}
\def\endproof{\hspace*{\fill}~\QED\par\endtrivlist\unskip}

\newcommand{\half}{\mbox{$\textstyle \frac{1}{2}$} }
\newcommand{\ket}[1]{| #1 \rangle}
\newcommand{\bra}[1]{\langle #1 |}
\newcommand{\proj}[1]{\ket{#1}\! \bra{#1}}
\newcommand{\outerp}[2]{\ket{#1}\! \bra{#2}}
\newcommand{\inner}[2]{ \langle #1 | #2 \rangle}
\newcommand{\melement}[2]{ \langle #1 | #2 | #1 \rangle}


\title{Quantum Rate-Distortion Coding}
\author{Howard Barnum}
\maketitle

\centerline{Hampshire College and ISIS}
\centerline{Amherst, MA 01002 USA}
\centerline{hbarnum@hampshire.edu}




\section{Introduction}
The fidelity criteria introduced in noisy and noiseless
coding theorems may seem excessively stringent.  The
classical criterion, for example, requires that the probability
of an error in the entire block approach zero as the block length
goes to infinity.  A code with a constant nonzero error 
rate per symbol would fail this test miserably (error probability 
would go to one in the large block limit), but could still be
perfectly acceptable as long as the error rate was sufficiently
small.  (Most, if not all, noisy channel coding protocols used
with real-world communications channels are examples.) 
Similarly in the quantum mechanical case, we might be
willing---taking an i.i.d. source for simplicity in the example---
to tolerate a constant rate of bad EPR pairs in the entanglement-
transmission case, or a finite deviation (``distortion'')
of the average pure
state fidelity of each transmission from one.  
A theory which tells us, given an ``error rate'' or level of
distortion which we have
decided we can tolerate, whether a given channel (noisy or noiseless)
can achieve that error rate, would be decidedly useful.  
This is rate-distortion theory.  

One
might think one could get by with substantially less resources
if one accepts the less ambititious fidelity criterion of requiring
a constant distortion rate.  
Classical rate-distortion theory 
tells us that there is no great savings
in allowing small average distortion rather than asymptotically 
perfect transmission.  Thus rate-distortion theory helps establish 
the relevance of theoretical
results like the asymptotic block-coding versions of noiseless and noisy
channel coding, to real-world schemes.  

\section{A quantum version of rate-distortion}

Let us use as our measure of distortion either one minus
the entanglement fidelity (for the entanglement transmission
problem) or one minus the average pure-state fidelity (for 
pure-state ensemble transmission problems).  This must be
evaluated for single transmissions, and then averaged over
the block of $n$ transmissions.  
I will confine myself to i.i.d. sources, with marginal density
operator $\rho$,  at least
initially.  Thus $\rho\pn \equiv \rho\on$.  
The channel will be taken to be noiseless;  then 
a $(n,2^{nR})$ {\em rate-distortion code} consists of a map ${\cal E}^{(n)}$ from 
n copies of the source space to n copies of 
a channel space of dimension $2^{nR}$,
followed by a decoding ${\cal D}^{(n)}$ from $n$ channels to 
$n$ source spaces.  The average distortion for an i.i.d. source
can then be 
defined as:
\begin{eqnarray}
D_e({\cal E}^{(n)},{\cal D}^{(n)}) \equiv \sum_{i=1}^n \frac{1}{n} (1-F_e (\rho, {\cal T}^{(n)}_i)),
\eeqa   
where ${\cal T}_i$ is the ``marginal operation''
on the $i$-th copy of the
source space induced by the overall operation ${\cal D}^{(n)}
\circ {\cal E}^n.$  More formally,
\beqa \label{eqtn: reducedoperation}
{\cal T}^{(n)}_i(\sigma) \equiv {\rm tr}_{Q_1,...Q_{i-1},
Q_{i+1},...,Q_{n}} [({\cal D}^{(n)}
\circ {\cal E}^n) (\rho \otimes \rho \cdots \otimes \rho \otimes
\sigma \otimes \rho \cdots \otimes \rho)],
\eeqa 
where the $\sigma$ in the input density operator is in the 
$i$-th position.  (It is easily checked that this defines
a tracepreserving operation.)
The same definition, but with $\overline{F}(E, {\cal T}^{(n)}_i)$,
as the fidelity criterion, defines the average pure-state
distortion $\overline{D}$.

$R$ is said to be the $rate$ of a rate-distortion code.  To avoid
confusion, I note here that the rate of a rate-distortion code
has a significance roughly inverse to that of 
the rate of information transmission
through a noisy channel.  (The terminology is already well-established
in classical information theory.)  The rate in rate-distortion is the
rate at which the source is described, that is, the number of qubits,
or the log of the
number of Hilbert space dimensions,  
used to encode the source, per source emission.  Thus the goal of
rate-distortion theory is to achieve low rates, i.e. to encode
the source into as few qubits as possible per source emission.

A rate-distortion pair $(R,D)$ is {\em achievable} for a given
source iff
there is a sequence of $(n,2^{nR})$ rate-distortion codes 
$({\cal E}^{(n)},{\cal D}^{(n)})$ such
that 
\beqa
\lim_{n \rightarrow \infty} D({\cal E}^{(n)},{\cal D}^{(n)})
\le D.
\eeqa

Here $D$ is whatever average distortion measure is used, e.g. 
$\overline{D}$ or $D_e$.
The {\em rate-distortion feasible set} for a source is
the closure of the set of achievable rate-distortion pairs.
The {\em rate-distortion function} $R(D)$ is defined by
\beqa
R(D) \equiv \inf{ R | (R,D) \mbox{is achievable}}.
\eeqa
The {\em rate-distortion frontier} is the graph of the rate
distortion function;  the {\em distortion-rate function} is
the inverse of the rate-distortion function.

If we assume that the coherent information continues to play the
role, in quantum information theory, of the mutual information in
classical information theory, then we are led to define quantum
analogues of the information rate-distortion function.

The {\em entanglement 
information rate-distortion function} $R^I(D)$ for
a source is defined by:
\beqa
R^I(D) \equiv \min_{{\cal A}| d({\cal A}) \le D} I_c(\rho,
{\cal A}).
\eeqa
One may conjecture that, as in the classical case, the information
rate-distortion function just defined is equal to the 
information-disturbance 
function defined above, and thus that $R^I(D)$ tells
us the lowest 
rate at which we can use channel qubits to end a quantum source with 
entanglement distortion no greater than $D$.  We might worry that
peculiarly quantum 
features such as the superadditivity of the coherent information
or the failure of data pipelining
require some modifications to the straightforward 
quantum analogue of the classical result, 
as they do in the case of noisy
channel coding.  In what follows, I
will derive
a lower bound on the required description rate;  perhaps this 
bound is not tight due to the peculiarly quantum effects just 
discussed, although the fact that general encodings are used in
deriving the bound makes me doubt that the failure of data 
pipelining is relevant.   I 
will not discuss achievability. I expect the techniques required 
for noisy channel coding may help in showing achievability, 
although rate-distortion
may be more difficult as we cannot rely on bounds that only become
tight for fidelities near one;  the saving grace may be that the
``noise''-like element is only truncation to a smaller space,
and this is likely to be much easier to deal with than a general
channel operation.

The proof I will give uses two lemmas.
First, we need the convexity of the information rate-distortion
function:

\begin{lemma} \label{lemma: ratedistconvexity}
$R^I(D)$ is a nonincreasing, convex function of $D$;  that is,
\beqa
D_1 < D_2 \rightarrow R^I(D_1) \ge R^I(D_2), \mbox{ and} \\
R^I(\lambda D_1 + (1-\lambda) D_2) \le 
\lambda R^I(D_1) + (1-\lambda) R^I(D_2)
\eeqa
where $0 \le \lambda \le 1$.
\end{lemma}

\begin{proof}
Nondecrease:  
As $D$ increases, the domain of the minimization
in the definition of $R^I(D)$ becomes larger (or at least no
smaller);  therefore, $R^I(D)$ does not increase.

Convexity:  Let $(R_1, D_1)$ and $(R_2,D_2)$ be points
on the information rate-distortion curve, and let ${\cal E}_1$ and
${\cal E}_2$ be operations achieving the minimum in the 
definition of $R^I(D)$ for $D= D_1$ and $D=D_2$ respectively.  
Consider the operation $\ce_\lambda \equiv
\lambda\ce_1 + 
(1-\lambda) \ce_2$.  Since the entanglement disturbance
is linear in the operation, this operation has disturbance
$D_\lambda \equiv D(\ce_\lambda) = \lambda D(\ce_1) + 
(1-\lambda) D(\ce_2).$  Since $R^I(D_\lambda)$ is the
minimum of the coherent information over operations,
$R^I(D_\lambda) \le I_c(\rho, \ce_\lambda).$
And since the coherent information is convex in the operation, 
this is less than $\lambda I_c(\rho, \ce_1) 
+ (1-\lambda) I_c(\rho, \ce_2) = 
\lambda R^I(D_1) + (1-\lambda) R^I(D_2)$.
\end{proof}

Notice that the only property of the disturbance that was
used in this proof was the linearity of the disturbance in
the operation;  hence it applies to any quantum rate-distortion
function defined using a disturbance measure
with this property, in particular to the information
rate-distortion function using average pure-state fidelity.

The second lemma we need is that the coherent information
for a process on a composite state is greater than or equal
to the total of the ``marginal coherent informations'' for 
the reductions of the process and the initial state to the
subsytems.  
\begin{lemma}\label{lemma: marginalcoherent}
$$
I_c(\rho\pn, \ce^{(n)})
\ge \sum_i I_c(\rho_i, {\cal E}_i^{(n)}).
$$
\end{lemma}
Here the definition of the reduced operation
${\cal E}_i^{(n)}$ is the same as that
of ${\cal T}^{(n)}_i$ in (\ref{eqtn: reducedoperation}), except
that ${\cal D}^{(n)}$ is omitted on the RHS.  $\rho_i$ is
of course the marginal density operator of the $i$-th system.
\begin{proof}
The lemma obviously follows from the two-system case:
\beqa
I_c(\rho^{(2)}, \ce^{(2)})
\ge I_c(\rho_1, {\cal E}_1^{(2)}) + I_c(\rho_2, \ce_2^{(2)}).
\eeqa
If we model this in the usual way, by purifying $Q_1$
into $R_1$ and $Q_2$ into $R_2,$ adjoining an initially
pure environment $E$ and effecting the operation 
$\ce^{(2)}$ by a unitary interaction $U^{Q_1 Q_2 E},$
this becomes:
\beqa
S(\rho^{Q_1 Q_2}) - S(\rho^{R_1 Q_1 R_2 Q_2})
\ge S(\rho^{Q_1}) + S(\rho^{Q_2}) - 
S(\rho^{R_1 Q_1}) - S(\rho^{R_2 Q_2})\;,
\eeqa
which may be rewritten
\beqa \label{eqtn: entanglementform}
S(\rho^{R_1 Q_1}) + S(\rho^{R_2 Q_2}) - S(\rho^{R_1 Q_1 R_2 Q_2})
\ge
S(\rho^{Q_1}) + S(\rho^{Q_2}) - S(\rho^{Q_1 Q_2}) \; .
\eeqa
The quantity appearing in this last form is the sum of 
the marginal entropies of two subsystems, 
minus the joint entropy of the composite system;  it is a 
quantity which can be larger in quantum theory than it
can in classical theory, due to entanglement.  In this form, 
the 
inequality says that this excess of marginal over joint
entropies is reduced if we ignore (trace over)
parts of each of the subsystems. This follows from strong 
subadditivity, as we may show by rewriting it yet again as:
\beqa \label{eqtn: yetanotherform}
S(\rho^{R_1 Q_1 R_2 Q_2}) + S(\rho^{Q_1}) + S(\rho^{Q_2})
\le
S(\rho^{Q_1 Q_2}) +
S(\rho^{R_1 Q_1}) + S(\rho^{R_2 Q_2})\;.
\eeqa
In this form, it follows from two applications of strong
subadditivity (thanks to Michael Nielsen for this observation).  
We start with a case of strong subadditivity for the
three systems $R_1$, $Q_1$, and $R_2Q_2$:
\beqa
S(\rho^{R_1 Q_1 R_2 Q_2}) + S(\rho^{Q_1}) \le
S(\rho^{R_1 Q_1}) + S(\rho^{Q_1 R_2 Q_2}) \;.
\eeqa
Adding $S(\rho^{Q_2})$ to both sides gives:
\beqa
S(\rho^{R_1 Q_1 R_2 Q_2}) + S(\rho^{Q_1}) + S(\rho^{Q_2}) \le
S(\rho^{R_1 Q_1}) + S(\rho^{Q_1 R_2 Q_2}) + S(\rho^{Q_2}) \;.
\eeqa
The last two terms on the right hand side are then upper bounded
by another application of strong subadditivity in the form
$S(\rho^{Q_1 R_2 Q_2}) + S(\rho^{Q_2}) \le S(\rho^{Q_1Q_2})
+ S(\rho^{R_2Q_2}),$ giving (\ref{eqtn: yetanotherform}).
\end{proof}

\begin{theorem} \label{theorem: ratedistortion bound}
Let $({\cal E}^{(n)},{\cal D}^{(n)})$ be a $(2^{nR}, n)$
rate-distortion code with distortion $D$.  Then
$R \ge R^I(D)$.
\end{theorem}

\begin{proof}
I give the proof
as a chain of inequalities and equivalences, 
followed by
notes justifying each inequality when possible.
\beqa
nR &\ge& S(\rho^{(n)'}) \label{dimstep}\\
&\ge& S(\rho^{(n)'}) - S_e(\rho, {\cal E}^{(n)})
\equiv I_c(\rho^{(n)},{\cal E}^{(n)})  \label{entexstep}\\
&\ge& I_c(\rho^{(n)},{\cal D}^{(n)}\circ{\cal E}^{(n)}) \label{dataprocstep}\\
&\ge& \sum_i  I_c(\rho_i,{\cal E}^{(n)}_i) \label{superaddstep}\\
&\ge& \sum_i R^I( d(\rho,{\cal E}^{(n)}_i)) \equiv 
n \sum_i \frac{1}{n} R^I( d(\rho,{\cal E}^{(n)}_i)) \label{ridefnstep}\\
&\ge& n R^I(\sum_i \frac{1}{n}d(\rho,{\cal E}^{(n)}_i))
\equiv n R^I(D).\label{convexitystep}
\eeqa
(\ref{dimstep}) holds because 
$nR$ is the log of the 
dimension of an $n$-block of channel Hilbert space, which constitutes
an uppper bound to the von Neumann entropy of a density operator on
that space.  (\ref{entexstep}) follows from the positivity of 
entropy exchange, (\ref{dataprocstep}) from the data processing
inequality, (\ref{superaddstep}) from Lemma 
\ref{lemma: marginalcoherent}, the superadditivity of 
coherent information compared to marginal coherent 
information, 
(\ref{ridefnstep}) follows from the definition of
the entanglement information rate-distortion function, and 
(\ref{convexitystep}) from Lemma (\ref{lemma: ratedistconvexity}),
the convexity of the rate-distortion
function.
\end{proof}






\end{document}